# Radiating jump conditions in General Relativity

**L.F. Castañeda-Godoy**[1,*], **J. Ospino** [2,†], **L.A. Núñez** [1,3,‡].

[1] Escuela de Física, Universidad Industrial de Santander, Bucaramanga-Colombia
[2] Departamento de Matemática Aplicada, Universidad de Salamanca, Salamanca-España.
[3] Departamento de Física, Universidad de los Andes, Mérida-Venezuela

E-mail: `Ludwin2198585@correo.uis.edu.co`[*], `j.ospino@usal.es`[†], `lnunez@uis.edu.co`[‡]

**Abstract.** We present a unified description of spherical discontinuity surfaces in General Relativity based on two parameters: mass function and surface permeability. The surfaces considered are: *Impulsive fronts*, massive permeable layer; *Surface layers*, massive impermeable layer; *Shock fronts*, massless permeable surface; and *Boundary surfaces*, massless impermeable surface. We derive the exact jump conditions for the physical variables across all these surfaces. Finally, we discuss the quasi-static approximation from studying slow hydrodynamic processes involving discontinuity surfaces.

## 1. Introduction

Discontinuity surfaces in Radiation Hydrodynamics have a long tradition of explosive events in both Newtonian and relativistic regimes. Supernova explosions, Gamma-ray bursts, accretion onto neutron stars or black holes usually involves modelling strong, radiant shock fronts. Some other not-so-violent discontinuities, like combustion fronts or deflagration waves, might be related to the phase transition of nuclear matter within cooling compact objects (see [1, 2, 3] and references therein).

The description of the dynamics of general relativistic radiant discontinuities is more elaborate and complicated because, in general, we might distinguish three different energy-momentum tensors: at the surface, ahead and behind it. Additionally, we have to fulfill junction conditions for the gravitational field generated by these energy-momentum tensors, i.e. study the continuity of the first and (dis)continuity of the second fundamental forms (see references [4], and [5] for an exciting discussion on relativistic junctions conditions).

This paper presents a unified description of spherical discontinuity surfaces in General Relativity based on two parameters: the Misner mass function [6] and the surface permeability.

This paper is organized as follows: in section 2 we describe the energy-momentum tensor and field equations for a non-static, anisotropic, spherically symmetric distribution of matter. Unified description for discontinuity surfaces is presented in section 3, while section 4 lists the particular jump conditions. The junction conditions under the quasi-static approximation are considered in section 5. Finally, we end with few conclusions in section 6.

## 2. Matter and radiation

Let us consider two spherically symmetric space-time with line elements denoted by subscripts $I$ and $II$ as

$$ds^2_{I,II} = e^{2\nu_{I,II}} dt^2 - e^{2\lambda_{I,II}} dr^2 - r^2(d\theta^2 + \sin^2\theta d\phi^2) \tag{1}$$

where the metric functions $\nu$ and $\lambda$ depend on $t$ and $r$.

The energy-momentum tensor for regions $I$ and $II$, $(T_{\mu\nu})_{I,II} = \left(T^M_{\mu\nu}\right)_{I,II} + \left(T^R_{\mu\nu}\right)_{I,II}$, have hydrodynamic and radiation counterparts, $\left(T^M_{\mu\nu}\right)_{I,II}$, and $\left(T^R_{\mu\nu}\right)_{I,II}$, respectively. Both can be written as

$$\left(T^M_{\mu\nu}\right)_{I,II} = (\rho + P_\perp) u_\mu u_\nu - P_\perp g_{\mu\nu} + (P - P_\perp) v_\mu v_\nu, \qquad \text{and} \tag{2}$$

$$\left(T^R_{\mu\nu}\right)_{I,II} = \frac{1}{2}(3\rho_R - \mathcal{P}) u_\mu u_\nu - \frac{1}{2}(\rho_R - \mathcal{P}) g_{\mu\nu} + \frac{1}{2}(\rho_R - 3\mathcal{P}) v_\mu v_\nu + \mathcal{F}_\mu u_\nu + \mathcal{F}_\nu u_\mu. \tag{3}$$

The variables are $\rho_H$, $P$ and $P_\perp$ represent the hydrodynamic density, radial and tangential pressures, respectively. The radiative contributions to the energy-momentum tensor are, respectively, the energy density $\rho_R$; the radiation pressure, $\mathcal{P}$, and $\mathcal{F}$, the energy flux.

Additionally

$$u_\mu = \gamma(e^\nu, -\omega e^\lambda, 0, 0), \quad v_\mu = \gamma(-\omega e^\nu, e^\lambda, 0, 0) \quad \text{and} \quad F_\mu = \mathcal{F} v_\mu, \tag{4}$$

with the Lorentz factor is $\gamma = \left(1 - \omega^2\right)^{-\frac{1}{2}}$ and with the fluid velocity defined as

$$\omega = \frac{dr}{dt} e^{\lambda - \nu} \tag{5}$$

The radiative contributions to the energy-momentum tensor are defined as the moments of specific intensity, $\mathbf{I}(r, t; \vec{n}, \nu)$, for a spherical (or Plano-paralelle) radiation field [1, 3]:

$$\rho_R = \frac{1}{2} \int_0^\infty d\nu \int_1^{-1} d\mu \, \mathbf{I}(r, t; \vec{n}, \nu), \quad \mathcal{F} = \frac{1}{2} \int_0^\infty d\nu \int_1^{-1} d\mu \, \mu \, \mathbf{I}(r, t; \vec{n}, \nu) \tag{6}$$

and

$$\mathcal{P} = \frac{1}{2} \int_0^\infty d\nu \int_1^{-1} d\mu \, \mu^2 \mathbf{I}(r, t; \vec{n}, \nu). \tag{7}$$

The corresponding Einstein equations $G^\nu_\mu = 8\pi T^\nu_\mu$ can be written in terms of the effective variables $\tilde{\rho} = \frac{\bar{\rho} + \bar{P}\omega^2 + 2\omega\mathcal{F}}{1 - \omega^2}$ and $\tilde{P} = \frac{\bar{P} + \bar{\rho}\omega^2 + 2\omega\mathcal{F}}{1 - \omega^2}$ as follows [7]:

$$m' = 4\pi r^2 \tilde{\rho}, \quad \nu' = \frac{4\pi r^3 \tilde{P} + m}{r(r - 2m)}, \quad \dot{m} = -\frac{4\pi r^2 e^{\nu - \lambda}}{1 - \omega^2} \left(\omega(\tilde{\rho} + \tilde{P}) + (1 + \omega^2)\mathcal{F}\right) \quad \text{and} \tag{8}$$

$$8\pi \bar{P}_\perp = \frac{e^{-2\nu}}{4} \left\{ 2\left(1 - \frac{2m}{r}\right)^{-1} \left(\frac{\ddot{m}}{r} + \frac{2\dot{m}^2}{r^2}\left(1 - \frac{2m}{r}\right)^{-1}\right) + \frac{\dot{m}}{r}\left(1 - \frac{2m}{r}\right)^{-1} \left(\frac{\dot{m}}{r}\left(1 - \frac{2m}{r}\right)^{-1} - \dot{\nu}\right) \right\}$$
$$+ \left\{ \left(1 - \frac{2m}{r}\right) \left(\nu'' + (\nu')^2 + \frac{\nu'}{r}\right) + \left(\frac{m}{r^2} - \frac{m'}{r}\right) \left(\nu' + \frac{1}{r}\right) \right\}, \tag{9}$$

where dots and primes represent derivatives with respect to $t$ and $r$, respectively.

The total physical variables (Hydro + Radiation) are defined as

$$\bar{\rho} = \rho + \rho_R, \quad \bar{P} = P + \mathcal{P} \quad \text{and} \quad \bar{P}_\perp = P_\perp + \mathcal{P}_\perp \quad \text{with} \quad \mathcal{P}_\perp = \frac{\rho_R - \mathcal{P}}{2}. \tag{10}$$

Finally, the metric function $\lambda(r, t)$ is expressed in terms of the Misner "mass function"[6] as

$$m(t, r) = \frac{r^2}{2} R^3_{232} \Leftrightarrow m(r, t) = 4\pi \int_0^r T^0_0 r^2 dr \Rightarrow e^{-2\lambda} = 1 - \frac{2m(r, t)}{r}. \tag{11}$$

## 3. Discontinuity surfaces in General Relativity

The dynamics of discontinuity surfaces are commonly used in General Relativity to describe various scenarios from supernova explosions to disc accretion onto a black hole [7, 8, 9, 10]. These relativistic surfaces are described through the junction conditions –i.e. the (dis)continuity of the first and second fundamental forms (see [4, 5] and reference therein)– and classified through the (dis)continuity of two parameters: the mass and the permeability. The jump in mass function leads to the discontinuity in the second fundamental form. The surface permeability is considered through the (dis)continuity of the matter velocity across the surface.

This classification of the relativistic discontinuity surfaces in terms of the mass and permeability of the surfaces is:

- **Impulsive fronts:** Massive permeable layer where the velocity of the front differs from the fluid velocities ahead and behind, i.e. $[m(r,t)]_c = m_{r=c_+} - m_{r=c_-} \neq 0$ and $\dot{c}e^{\lambda_c - \nu_c} \equiv \frac{dc(t)}{dt}e^{\lambda_c - \nu_c} \neq \omega_{c_-} \neq \omega_{c_+}$, with a discontinuous second fundamental;
- **Surface layers:** Massive impermeable layer where the velocity of the front coincides with the fluid velocities ahead and behind, i.e. $[m(r,t)]_c = m_{r=c_+} - m_{r=c_-} \neq 0$ and $\dot{c}e^{\lambda_c - \nu_c} \equiv \frac{dc(t)}{dt}e^{\lambda_c - \nu_c} = \omega_{c_-} = \omega_{c_+}$, with a discontinuous second fundamental;
- **Shock fronts:** Massless permeable surface where the velocity of the front differs from the fluid velocities ahead and behind, i.e. $[m(r,t)]_c = m_{r=c_+} - m_{r=c_-} = 0$ and $\dot{c}e^{\lambda_c - \nu_c} \equiv \frac{dc(t)}{dt}e^{\lambda_c - \nu_c} \neq \omega_{c_-} \neq \omega_{c_+}$ with a continuous second fundamental;
- **Boundary surfaces:** Massless impermeable surface where the velocity of the front coincides with the fluid velocities ahead and behind, i.e. $[m(r,t)]_c = m_{r=c_+} - m_{r=c_-} = 0$ and $\dot{c}e^{\lambda_c - \nu_c} \equiv \frac{dc(t)}{dt}e^{\lambda_c - \nu_c} = \omega_{c_-} = \omega_{c_+}$ with a continuous second fundamental.

The subscripts $c$, $c_+$ and $c_-$, indicate that the variable is evaluated at, ahead and behind the surface, respectively.

## 4. Junction and jump conditions

The first fundamental form is continuous for all of the above types of discontinuity surfaces and can be written as

$$\left[e^{2\nu}\left(1 - \omega^2\right)\right]_c = 0. \tag{12}$$

Our classification derives from the (dis)continuity across the surface of two parameters: the mass function (second fundamental form) and permeability (fluid velocity).

To evaluate the consequences of the (dis)continuity of the second fundamental form, we followed the Herrera-Jimenez scheme for junction conditions based on the Newman-Penrose formalism [11]. The metric (1) is described in terms of a complex null tetrad as $g_{\mu\nu} = n_\mu l_\nu + l_\mu n_\nu - m_\mu \bar{m}_\nu - \bar{m}_\mu m_\nu$, with the corresponding null vectors written as:

$$l_\mu = \frac{1}{\sqrt{2}}e^\nu \delta_\mu^0 + \frac{1}{\sqrt{2}}e^\lambda \delta_\mu^1, \quad n_\mu = \frac{1}{\sqrt{2}}e^\nu \delta_\mu^0 - \frac{1}{\sqrt{2}}e^\lambda \delta_\mu^1, \quad \text{and} \quad m_\mu = \frac{r}{\sqrt{2}}\delta_\mu^2 + \frac{i}{\sqrt{2}}r\sin\theta \delta_\mu^3,$$

and the significant non-vanishing spin coefficients are

$$\gamma = -\frac{1}{2\sqrt{2}}\left(\dot{\lambda}e^{-\nu} + \nu' e^{-\lambda}\right) \quad \text{and} \quad \epsilon = \frac{1}{2\sqrt{2}}\left(\dot{\lambda}e^{-\nu} - \nu' e^{-\lambda}\right), \tag{13}$$

see [12, 13, 14] and references therein.

The equations for the discontinuities of the physical variables across the different types of surfaces described above are:

- **Impulsive fronts:** The generalized Rankine-Hugoniot conditions for impulsive shock [13] are found from the discontinuity of the spin coefficients ($\gamma$, $\epsilon$) and using the Einstein's equations (8)

$$\dot{c}\left[\dot{m}\left(1-\frac{2m}{r}\right)^{-3/2}e^{-2\nu}\right]_c + \left[\frac{4\pi r^3 \tilde{P}+m}{r-2m}\left(1-\frac{2m}{r}\right)^{1/2}\right]_c =$$
$$\sqrt{2}c\dot{c}\left[e^{-2\nu}\left(1-\frac{2m}{r}\right)^{-1/2}(\epsilon-\gamma)\right]_c - \sqrt{2}c\left[\gamma+\epsilon\right]_c, \quad \text{and} \qquad (14)$$

$$\left[\dot{m}+4\pi r^2 \dot{c}\tilde{\rho}\right]_c - \{[m]_c\}^{\cdot} = 0. \qquad (15)$$

- **Surface layers:** Assuming impermeability of the surface, $\dot{c}e^{\lambda_c-\nu_c} = \omega_{c_-} = \omega_{c_+}$, in equation (14) we obtain the jump condition for the physical variables across a surface layer [13, 14]:

$$4\pi c^2 \left[\left(\bar{P}+\mathcal{F}\omega\right)\left(1-\frac{2m}{r}\right)^{-1/2}\right]_c + \left[\frac{m}{r}\left(1-\frac{2m}{r}\right)^{-1/2}\right]_c = -\sqrt{2}c\left([\gamma]_c(\omega-1)-[\epsilon]_c(\omega+1)\right). \qquad (16)$$

The condition, (15), associated with the mass function's discontinuity remains the same.

- **Shock fronts:** Now, assuming a massless surface in the generalized Rankine-Hugoniot conditions (14), we obtain

$$\left[\frac{\dot{c}\left(1-\frac{2m}{r}\right)^{-1/2}e^{-\nu}}{1+\omega^2}\left(\omega\left(\tilde{\rho}+\tilde{P}\right)+\left(1-\omega^2\right)\mathcal{F}\right)-\tilde{P}\right]_c = 0 \quad \text{and} \qquad (17)$$

$$\left[\dot{c}\tilde{\rho}-\frac{e^{\nu-\lambda}}{1+\omega^2}\left(\omega(\tilde{\rho}+\tilde{P})+(1-\omega^2)\mathcal{F}\right)\right]_c = 0. \qquad (18)$$

These are the well known general relativistic Rankine-Hugoniot conditions[15, 16, 17, 8, 9, 7].

- **Boundary surfaces:** Again, assuming impermeable surfaces, i.e. $\dot{c}e^{\lambda_c-\nu_c} = \omega_{c_-} = \omega_{c_+}$, in equations (17) and (18), we obtain the jump conditions for boundary surfaces[4, 5]:

$$-\omega\left[\mathcal{F}\right]_c = \left[\bar{P}\right]_c \qquad \text{and} \qquad -\left[\mathcal{F}\right]_c = \omega\left[\bar{P}\right]_c. \qquad (19)$$

These equations are exact for spherical (and Plano-parallel) matter configurations and easily approximated for particular extreme radiation-hydrodynamic scenarios.

In the case of radiation dominated environment we assume $T_{\mu\nu} \approx T^R_{\mu\nu}$ [18] and the physical variables become:

$$\bar{\rho} \approx \rho_R, \quad \bar{P} \approx \mathcal{P}, \quad \tilde{\rho} \approx \frac{\rho_R + \mathcal{P}\omega^2 + 2\omega\mathcal{F}}{1-\omega^2} \quad \text{and} \quad \tilde{P} = \frac{\mathcal{P}+\rho_R\omega^2+2\omega\mathcal{F}}{1-\omega^2}. \qquad (20)$$

On the other hand, for hydrodynamic dominated conditions, i.e. $T_{\mu\nu} \approx T^M_{\mu\nu}$, the matter variables are:

$$\bar{\rho} \approx \rho, \quad \bar{P} \approx P, \quad \tilde{\rho} \approx \frac{\rho+P\omega^2+2\omega\mathcal{F}}{1-\omega^2} \quad \text{and} \quad \tilde{P} = \frac{P+\rho\omega^2+2\omega\mathcal{F}}{1-\omega^2}. \qquad (21)$$

## 5. Quasi-static approximation

To gain some physical insight for the different type of discontinuity surfaces we will consider the quasi-static fluid evolution, assuming $\omega^2 \approx \ddot{\lambda} \approx \dot{\lambda}^2 \approx \dot{\lambda}\dot{\nu} \approx \ddot{\nu} \approx 0$. The quasi-static evolution approximation means that changes of the system take place in a very long time-frame compared with the hydrostatic time scale. Thus, the physical variables are functions of time but the system can be considered in hydrostatic equilibrium (see [19, 20, 21] and references therein). This approximation could be applied to model slow combustion waves, weak shocks, slim discontinuities and other slow jump surfaces.

The equations for the discontinuities of the physical variables for the different type of surfaces described above are:

- **Quasi-static impulsive shock fronts:**

$$\left[\bar{P}\left(1-\frac{2m}{r}\right)^{-1/2}\right]_c + \frac{1}{4\pi c^3}\left[m\left(1-\frac{2m}{r}\right)^{-1/2}\right]_c = \frac{\sqrt{2}\dot{c}}{4\pi c}\left[e^{-2\nu}\left(1-\frac{2m}{r}\right)^{-1/2}(\epsilon-\gamma)\right]_c - \frac{\sqrt{2}}{4\pi c}\left([\gamma+\epsilon]_c\right), \quad (22)$$

$$\left[\left(\dot{c}-\omega e^\nu\left(1-\frac{2m}{r}\right)^{1/2}\right)\bar{\rho}\right]_c - \left[\omega e^\nu\left(1-\frac{2m}{r}\right)^{1/2}\bar{P}\right]_c - \left[\mathcal{F}e^\nu\left(1-\frac{2m}{r}\right)^{1/2}\right]_c = \frac{1}{4\pi c^2}\{[m]_c\}^{\cdot}. \quad (23)$$

- **Quasi-static surface layers:**

$$\left[\bar{P}\left(1-\frac{2m}{r}\right)^{-1/2}\right]_c + \frac{1}{4\pi c^3}\left[m\left(1-\frac{2m}{r}\right)^{-1/2}\right]_c = \frac{\sqrt{2}\omega}{4\pi c}\left[e^{-\nu}(\epsilon-\gamma)\right]_c - \frac{\sqrt{2}}{4\pi c}\left([\gamma+\epsilon]_c\right), \quad (24)$$

$$-\omega\left[e^\nu\left(1-\frac{2m}{r}\right)^{1/2}\bar{P}\right]_c - \left[\mathcal{F}e^\nu\left(1-\frac{2m}{r}\right)^{1/2}\right]_c = \frac{1}{4\pi c^2}\{[m]_c\}^{\cdot}. \quad (25)$$

- **Quasi-static shock fronts:**

$$[\bar{P}]_c = 0 \quad \text{and} \quad \left[\left(\dot{c}-\omega e^\nu\left(1-\frac{2m}{r}\right)^{1/2}\right)\bar{\rho}\right]_c - \left[\omega e^\nu\left(1-\frac{2m}{r}\right)^{1/2}\bar{P}\right]_c - \left[\mathcal{F}e^\nu\left(1-\frac{2m}{r}\right)^{1/2}\right]_c = 0. \quad (26)$$

- **Quasi-static boundary surfaces:**

$$[\bar{P}]_c = 0 \quad \text{and} \quad [\mathcal{F}]_c = 0. \quad (27)$$

Again, in equations (22)-(23), (24)-(25), (26) and (27) the above limits can be considered –i.e. radiation-dominated $T_{\mu\nu} \approx T_{\mu\nu}^R$ and matter-dominated $T_{\mu\nu} \approx T_{\mu\nu}^M$– through a change of variables (20) and (21) are applied.

## 6. Conclusions

We presented a unified description of spherical discontinuity surfaces in General Relativity based on two parameters: mass function and surface permeability. The most general discontinuities are impulsive shocks (massive impermeable surfaces). The other surfaces can be deduced from these general jump conditions, tuning off the surface's mass or/and making it permeable. In all the cases, the evolution of the surface, $\dot{c}$ can be solved and depends on the discontinuities of physical variables.

We applied the quasi-static approximation to present the jump conditions for other slow physical phenomena (slow combustion waves, weak shocks, slim discontinuities and other slow jump surfaces).

Both set of equations (14) through (19) are exact for the spherical case and can be easily applied to extreme scenarios such as radiation-dominated $T_{\mu\nu} \approx T_{\mu\nu}^R$ or matter-dominated $T_{\mu\nu} \approx T_{\mu\nu}^M$, implementing a simple changes of variables.

## 7. Acknowledgments

We gratefully acknowledge the financial support of the Vicerrectoría de Investigación y Extensión of the Universidad Industrial de Santander and COLCIENCIAS under project No. 8863. J.O. thanks Ministerio de Ciencia, Innovacion y Universidades, Spain for Grant PGC 2018-096038-B-I00, and Junta de Castilla y Leon, Spain, grant number: SA096P20.